\documentstyle[epsf,aaspp4]{article}

\newcommand{\simgt}{\lower.5ex\hbox{$\; \buildrel > \over \sim \;$}}
\newcommand{\simlt}{\lower.5ex\hbox{$\; \buildrel < \over \sim \;$}}
\newcommand{\citet}[1] {\cite{#1}}
\newcommand{\citep}[1] {(\cite{#1})}
 \newcommand{\bm}[1]{\mbox{\boldmath$#1$}}
 \newcommand{\kaco}[1]{\left\langle{#1}\right\rangle}
 
\begin{document}
\title{Gravitational Lensing Effect on the Two-point
Correlation of Hotspots in the Cosmic Microwave Background}


\author{Masahiro Takada\altaffilmark{1}, Eiichiro Komatsu\altaffilmark{1,2}
and Toshifumi Futamase\altaffilmark{1}}

\altaffiltext{1}{Astronomical Institute, Graduate School of Science, 
Tohoku University, Sendai 980-8578, Japan; takada@astr.tohoku.ac.jp,
komatsu@astr.tohoku.ac.jp, tof@astr.tohoku.ac.jp}
\altaffiltext{2}{Department of Astrophysical Sciences, Princeton
University, Princeton, NJ 08544, USA; komatsu@astro.princeton.edu}

\bigskip
\begin{abstract}
We investigate the weak gravitational lensing effect due to the
 large-scale structure of the universe on two-point correlations 
of local maxima ({\em hotspots}) in the 2D sky map of the cosmic
microwave background (CMB) anisotropy. According to the Gaussian random
statistics as most inflationary scenarios predict, 
the hotspots are discretely distributed with some {\em characteristic}
angular separations on the last scattering surface owing to
oscillations of the CMB angular power spectrum.
 The weak lensing then causes pairs of hotspots which are separated
with the characteristic scale to be observed with various separations. 
We found that the lensing fairly smoothes
 the oscillatory features of the two-point correlation function of
hotspots. This indicates that the hotspots correlations can be a new
statistical tool for measuring shape and normalization of
the power spectrum of matter fluctuations from the lensing signatures.
\end{abstract}

\section{Introduction}
The temperature anisotropy of the cosmic microwave background (CMB)
is the most powerful probe of our universe because the anisotropies
encode a wealth of cosmological information and the physical processes
involved in anisotropy formation are well understood
by linear perturbation theory (e.g., \cite{HSS}). Therefore, high-precision
measurements of the primary CMB anisotropies enable us to determine a
lot of fundamental cosmological parameters with unprecedented precision
(\cite{Jungman}; \cite{Bond}; \cite{Zald97}).
On the other hand, gravitational lensing due to the
large-scale structure between the last scattering surface (LSS)
and us deforms the 2D sky map of the CMB anisotropy.
The weak lensing will be in the near future an effective tool for
 mapping the inhomogeneous distributions of dark matter
because it is not affected by uncertainties in the ``biasing'' problem
regarding the extent to which luminous objects trace the mass
distribution. Furthermore, a great advantage in studying the lensing effect
on the CMB anisotropies is that statistical properties of the unlensed CMB field
can be fully specified once we assume the field obeys the Gaussian random
theory (\cite{BBKS}, hereafter BBKS; \cite{BE} hereafter BE). Most 
inflationary scenarios indeed support this assumption. However,
most previous studies have focused on theoretical predictions of lensed shapes
of the two-point correlation function $C(\theta)$ or equivalently the
angular power spectrum $C_l$ of the temperature fluctuations
field (\cite{Lens}; \cite{Cole}; \cite{Sasaki}; \cite{Tomita};
Cay\'on, Mart\'nez-Gonz\'alez \& Sanz 1993a, b; \cite{Seljak96};
\cite{Metcalf97}; \cite{ZS}), while some studies have shown that
the lensing induces non-Gaussian signatures (\cite{Ber97}; \cite{Zald}).
It is therefore worth exploring an other statistical CMB estimator sensitive
to the lensing signatures using more directly the advantage of predictability of
the 2D CMB sky map.  

Recently, Heavens and Sheth (1999, hereafter HS) investigated 
the two-point correlation function of local maxima ({\em hotspots})
above a certain threshold in the 2D CMB map, say $\xi_{\rm pk-pk}(\theta)$, 
which  can be accurately calculated based on the Gaussian random
theory once $C_l$ is given. Their purpose was to propose that $\xi_{\rm
pk-pk}(\theta)$ can be a powerful test of the Gaussian hypothesis of the 
temperature fluctuations, and they showed that the shape of $\xi_{\rm
pk-pk}(\theta)$ is largely different from that of $C(\theta)$.
$\xi_{\rm pk-pk}(\theta)$ has oscillatory
features such as the depression around the angular scale corresponding to
the 1st Doppler peak of $C_l$ and a prominent peak with a damping tail
toward the smaller scales than the peak-scale. The reasons
can be explained as follows. In the derivation
of $\xi_{\rm pk-pk}(\theta)$, we need statistical properties of the
gradient and second derivative fields of
the temperature fluctuations field in order to identify the local maxima 
in the 2D CMB map. The shape of $\xi_{\rm pk-pk}(\theta)$ thus
reflects more strongly the oscillatory features of $C_l$ such as
a series of Doppler peaks than $C(\theta)$ does, because
the gradient and the second derivative fields are more sensitive to the
oscillations of $C_l$.
The oscillations of $\xi_{\rm pk-pk}(\theta)$ physically mean 
that the hotspots are discretely distributed with some
{\em characteristic} angular separations on the LSS. Then, the weak
lensing due to the large-scale structure causes the pairs of hotspots
separated with the characteristic scale on the LSS to be observed
with various separations in the random line of
sight. This leads to the expectation that $\xi_{\rm pk-pk}(\theta)$
enables us to see more directly the lensing dispersion of the specific angular
separation than using $C(\theta)$. The purpose of this Letter
is therefore to develop the formula for calculating the lensing effect on
$\xi_{\rm pk-pk}(\theta)$ using
the power spectrum approach to compute the dispersion of the lensing
deflection angle (\cite{Seljak96}), and to present some results in 
the standard cold dark matter (SCDM) model. 

\section{Weak Lensing Effect on the Hotspots Correlation Function}
During the propagation from the LSS to us,
a CMB photon is randomly deflected by the inhomogeneous matter
distributions in the intervening large-scale structure. 
Then, two CMB photons observed with a certain angular separation $\theta$
have  a different separation when emitted on the LSS.
If we assume that
the lensing fluctuations of the relative angular separation,
$\delta\bm{\theta}(=\delta\bm{\theta}_1
-\delta\bm{\theta}_2)$, obey the 
Gaussian random statistics (\cite{Seljak96}), 
the ensemble average of the following characteristic function can be
calculated as 
\begin{equation}
\kaco{\exp[i\bm{l}\cdot(\delta\bm{\theta}_1-\delta\bm{\theta}_2)]}
_{|\bm{\theta}_1 -\bm{\theta}_2|=\theta}\simeq 
\exp\left[-\frac{l^2}{2}\sigma^2_{\rm GL}(\theta)\right],
\label{eqn:avergl}
\end{equation}
where $\delta \bm{\theta}_1$ and $\delta \bm{\theta}_2$ are the angular
excursions of the two photons, and $\kaco{\ \ }_\theta$ denotes the
average performed over all pairs with a fixed observed angular
separation $\theta$. The $\sigma_{\rm GL}^2(\theta)$ is defined by
\begin{eqnarray}
&&\sigma_{\rm GL}^2(\theta)\!=\!\frac{9}{2\pi}H_0^4\Omega_{\rm m0}^2
\int^{\infty}_0\!\!\frac{dk}{k}\int^{\chi_{\rm rec}}_0\!\!\!d\chi 
a^{-2}P_{\delta}(k,\tau)W^2\!(\chi,\chi_{\rm rec})
(1\!-\!J_0(\chi k\theta)), \label{eqn:gldisp}
\end{eqnarray}
where $\tau$ is a conformal time, $\chi\equiv \tau_0-\tau$,
$J_n(x)$ is the Bessel function of order $n$, and 
the subscript $0$ and ``rec'' denote values at present and a
recombination time, respectively. 
$P_\delta(k,\tau)$ is the power spectrum of the matter
fluctuations, $H_0=100h$ km s$^{-1}$ Mpc$^{-1}$ and $\Omega_{\rm m0}$ denote
the present Hubble parameter
and the present energy density of matter, respectively. 
The projection operator $W(\chi,\chi_{\rm rec})$ on the celestial sphere
is given by $W(\chi,\chi_{\rm rec})=1-\chi/\chi_{\rm rec}$
in a flat universe. The approximation used in the derivation of equation 
(\ref{eqn:avergl}) is valid as
long as the $\sigma_{{\rm GL}}(\theta)/\theta\ll 1$ is satisfied, and  
we numerically  confirmed $\sigma_{{\rm GL}}/\theta<0.25$ on the relevant 
angular scales for all normalizations of matter power spectrum
considered in this Letter.  

The unlensed $C(\theta)$
in the context of the small angle approximation developed by BE
 is given by
$
C(\theta)=\kaco{\Delta(\bm{\theta}_1)\Delta(\bm{\theta}_2)}=\int
(ldl/2\pi)S(l)J_0(l\theta),
$
where $|\bm{\theta}_1-\bm{\theta}_2|=\theta$ and
$\Delta(\bm{\theta})\equiv (T(\bm{\theta})-T_{\rm CMB})/T_{\rm CMB}$.
The 2D power spectrum of the temperature fluctuations field, 
$S(l)$, is identical to $C_l$ in the limit of $l\gg 1$.

First, we briefly present the derivation
of the unlensed two-point correlation function of hotspots
in the 2D Gaussian random CMB field developed by HS.
The moments of power spectrum are defined by
$
\sigma_i^2\equiv \int^\infty_0 (ldl/2\pi)S(l)l^{2i}.
$
At a hotspot, the gradients of temperature fluctuations field, 
$\Delta_i\equiv\partial \Delta/\partial \theta^i$, vanish 
and the eigenvalues of the second derivative matrix, 
$\Delta_{ij}\equiv \partial^2\Delta/\partial \theta^i\partial 
\theta^j$, are negative. We thus need 6 independent variables $\bm{v}
=(\Delta,
\Delta_x,\Delta_y,\Delta_{xx},\Delta_{xy},\Delta_{yy})$ 
to specify one local hotspot.  The probability density function
of the 12 variables $\bm{v}\equiv(\bm{v}_{(1)},\bm{v}_{(2)})
=(\bm{v}(\bm{\theta}_1),\bm{v}(\bm{\theta}_2))$
 for the two hotspots separated with the angle
 $\theta(=|\bm{\theta}_1-\bm{\theta}_2|)$ is defined by
\begin{equation}
p_{2}(\bm{v}_{(1)},\bm{v}_{(2)})=\frac{1}{(2\pi)^6|M|^{1/2}}
\exp{\left[-\frac{1}{2}v_i M^{-1}_{ij}v_j\right]},
\end{equation}
where $M_{ij}$ is the covariance matrix: $M_{ij}\equiv
\kaco{(v_i-\kaco{v_i})(v_j-\kaco{v_j})}$, 
$M={\rm det}(M_{ij})$, and $M^{-1}_{ij}$ is the 
inverse of $M_{ij}$. Note that $\kaco{v_i}=0$ in the present case.
It is then convenient to introduce the variables defined by
%
$
\nu\equiv\Delta/\sigma_0,
\eta_i\equiv\Delta_i/\sigma_1,
X\equiv-(\Delta_{xx}+\Delta_{yy})/\sigma_2,
Y\equiv(\Delta_{xx}-\Delta_{yy})/\sigma_2,$ and $
Z\equiv2\Delta_{xy}/\sigma_2.
$
%
All non-zero components of $M_{ij}$ for two hotspots characterized 
by $\nu_{(1)}\equiv \nu(\bm{\theta}_1)$ and $\nu_{(2)}\equiv\nu(\bm{\theta}_2)$
and so on were explicitly calculated by HS. 
Employing the BBKS formalism, we can obtain the following
unlensed correlation function of two hotspots above certain thresholds
$\nu_1$ and $\nu_2$, respectively:
\begin{eqnarray}
   \nonumber 
1 + \xi_{\rm pk-pk}(\theta|>\!\nu_1,>\!\nu_2)&\equiv&
\frac{1}{\bar{n}_{\rm pk}(>\nu_1)\bar{n}_{\rm pk}(>\nu_2)}
\kaco{n_{\rm pk}(\bm{\theta}_1)n_{\rm pk}(\bm{\theta}_2)}\nonumber \\
&&\hspace{-7em}=
   \frac{1}
   {2^2\theta_\ast^4\bar{n}_{\rm pk}(>\nu_1)\bar{n}_{\rm pk}(>\nu_2)}
   \int^\infty_{\nu_1}\!d\nu'_{(1)}\!
   \int^\infty_{\nu_2}\!d\nu'_{(2)}\!
   \int^\infty_{0}\!dX_{(1)}\!
   \int^\infty_{0}\!dX_{(2)}\!
   \int^{X_{(1)}}
       _{-X_{(1)}}\!dY_{(1)}\!
   \int^{X_{(2)}}
       _{-X_{(2)}}\!dY_{(2)}\!\nonumber \\
&&\hspace{-5em}
   \times \ \int^{\sqrt{X_{(1)}^2-Y_{(1)}^2}}
       _{-\sqrt{X_{(1)}^2-Y_{(1)}^2}}\!dZ_{(1)}\!
   \int^{\sqrt{X_{(2)}^2-Y_{(2)}^2}}_{-\sqrt{X_{(2)}^2-Y_{(2)}^2}}\!dZ_{(2)}
   \label{eqn:xi}
       (X_{(1)}^2-Y^2_{(1)}-Z_{(1)}^2)
       (X_{(2)}^2-Y^2_{(2)}-Z_{(2)}^2)
\nonumber\\
&&\hspace{-5em}
       \times \ p_2\!(
	       \nu'_{(1)},
	       X_{(1)},Y_{(1)},Z_{(1)},
	       \eta_{(1)i}=0,
	       \nu'_{(2)},
	       X_{(2)},Y_{(2)},Z_{(2)},
	       \eta_{(2)i}=0
	      ),
\end{eqnarray}
where $\theta_{\ast}=\sqrt{2}\sigma_1/\sigma_2$,
$n_{\rm pk}(\bm{\theta})$ and $\bar{n}_{\rm pk}(>\!\nu)$
are the number density filed and the mean number density of
hotspots above height $\nu$, respectively. To obtain $\xi_{\rm
pk-pk}(\theta)$, we performed a 6D numerical integration of
equation (\ref{eqn:xi}) because the two can be done analytically. 

Next, we derive the lensed correlation function of hotspots. 
In this Letter, we focus on the lensing contributions to $\xi_{\rm
pk-pk}(\theta)$ as a secondary effect which comes from a
mapping of the intrinsic discrete distributions of hotspots on the LSS.
The procedure is appropriate because a power of anisotropies generated
by the lensing is small (\cite{SZ99}; \cite{Zald}).
For this purpose, 
we define the fluctuations field of $n_{\rm pk}(\bm{\theta})$ as
$\delta n(\bm{\theta})\equiv (n_{\rm
pk}(\bm{\theta})-\bar{n}_{\rm pk})/\bar{n}_{\rm pk}$.
Likewise, since the gravitational lensing makes us to observe, at a
certain angular position $\bm{\theta}$, a hotspot at a different position 
$\bm{\theta}+\delta\bm{\theta}$ on the LSS, the lensed (observed)
fluctuations field $\delta\tilde{n}(\bm{\theta})$ can be expressed as
$
\delta\tilde{n}(\bm{\theta})\equiv \delta n(\bm{\theta}
+\delta\bm{\theta})=\int\!d^2\bm{l}/(2\pi)^2\delta n(\bm{l})
\exp[i\bm{l}\cdot(\bm{\theta}+\delta\bm{\theta})],
$
where $\delta n(\bm{l})$ is the Fourier component of $
\delta n(\bm{\theta})$. If assuming that the lensing 
fluctuations and the intrinsic CMB anisotropies field are statistically
independent, then the lensed two-point correlation function of hotspots
can be calculated with help of equation 
(\ref{eqn:avergl}) as
\begin{equation}
\xi^{\rm GL}_{\rm pk-pk}(\theta)\equiv\kaco{\delta\tilde{n}
 (\bm{\theta}_1)\delta\tilde{n}(\bm{\theta}_2)}
\simeq\frac{1}{\sigma^2_{\rm GL}(\theta)}\int\!\!d\theta'\theta'
\xi_{\rm pk-pk}(\theta')
\exp\left[-\frac{\theta^2+\theta'^2}{2\sigma_{\rm GL}^2(\theta)}\right]
I_0\!\left(\frac{\theta\theta'}{\sigma_{\rm GL}^2(\theta)}\right).
\label{eqn:lensxi}
\end{equation}
where $I_0(x)$ is the modified zeroth-order Bessel function. 
This equation indicates that
the lensing contribution at a certain scale $\theta$ arises only from the
lensing dispersion $\sigma_{\rm GL}(\theta)$ at the same scale alone.
 Hence detecting the scale dependence of the lensing
contributions to $\xi_{\rm pk-pk}(\theta)$ enables us to reconstruct the
lensing dispersion, more importantly the projected matter power spectrum by
equation (\ref{eqn:gldisp}). 

\section{Results}
In the following, we show some results in the
SCDM model with $\Omega_{\rm m0}=1$. 
The Hubble parameter and the present baryon density
are $h=0.6$ and $\Omega_{\rm b0}h^2=0.015$, respectively. 
To compute the intrinsic CMB angular power spectrum, $S(l)$, 
we used the CMBFAST code developed by Seljak \& Zaldarriaga (1996).
As for the matter power spectrum in the lensing dispersion (\ref{eqn:gldisp}),
 we employed  the Harrison-Zel'dovich spectrum and
the BBKS transfer function with the shape parameter in
Sugiyama (1995).
A free parameter in our model is only the normalization of the present-day
matter power spectrum which is conventionally expressed in terms of the
rms mass fluctuations within a sphere of $8h^{-1}$Mpc, i.e.,
$\sigma_8$. 

\begin{figure}
 \begin{center}
  \leavevmode\epsfysize=17.0cm \epsfbox{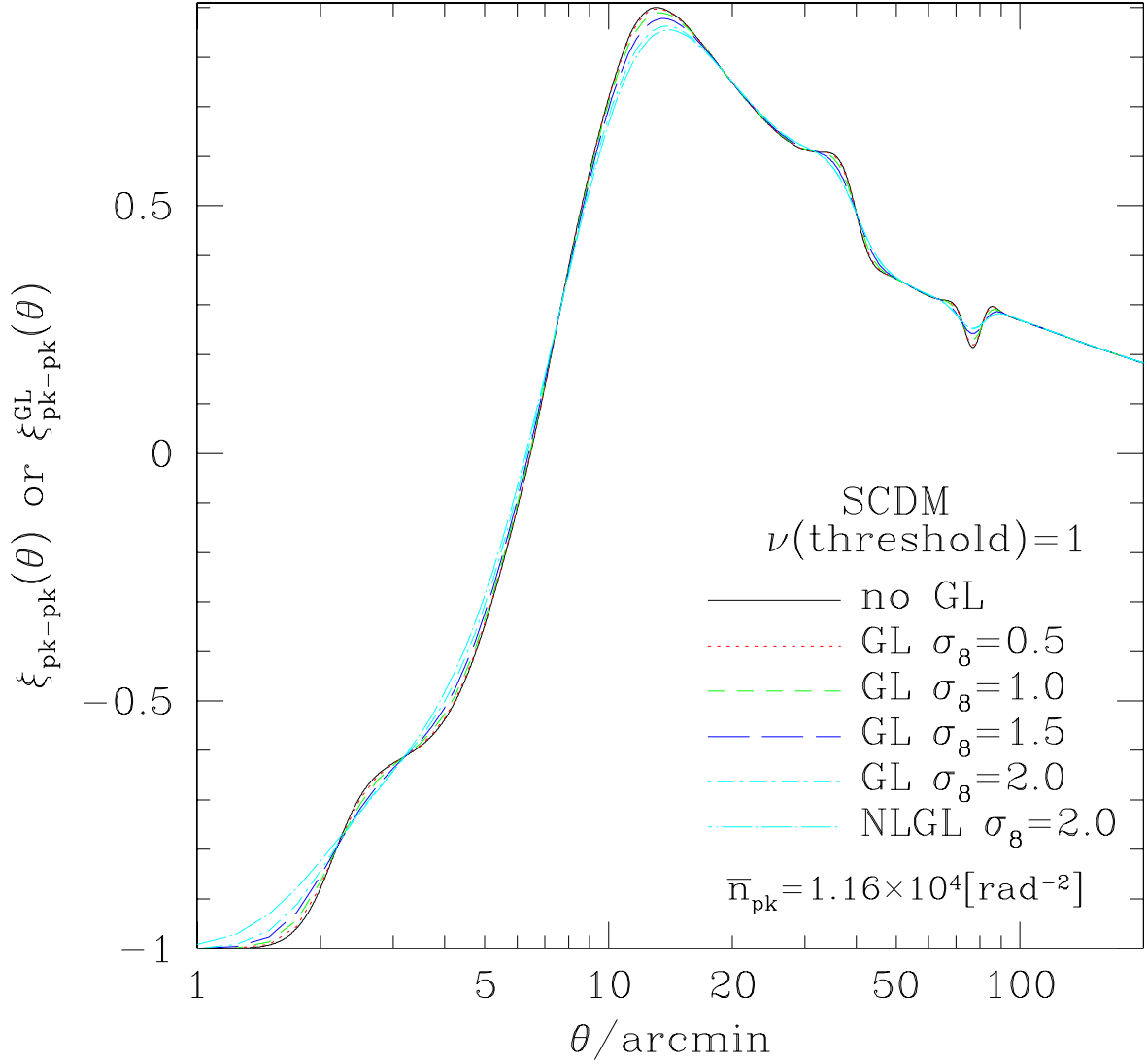}
  \end{center}
\figcaption{The correlation function of hotspots 
  vs. $\theta$ with  
  and without the gravitational lensing effect in SCDM model.
  The solid line is the unlensed correlation function, while dotted,
  short-dashed, long-dashed and dotted-dashed lines represent the corresponding
  lensing cases with $\sigma_8=0.5,1.0,1.5$ and $2.0$, respectively. The
  long-dotted-dashed line denotes the result including nonlinear correction with 
  $\sigma_8=2.0$ (see text). $h=0.6$ and
  $\Omega_{\rm b0}h^2=0.015$ are assumed. The threshold of hotspots is
  $\nu=1$. The mean density of hotspots is then $\bar{n}_{\rm
  pk}=1.16\times 10^4 {\rm rad}^{-2}$. 
 \label{fig:scdmxi} }
 \end{figure}

\begin{figure}
 \begin{center}
  \leavevmode\epsfysize=8.0cm \epsfbox{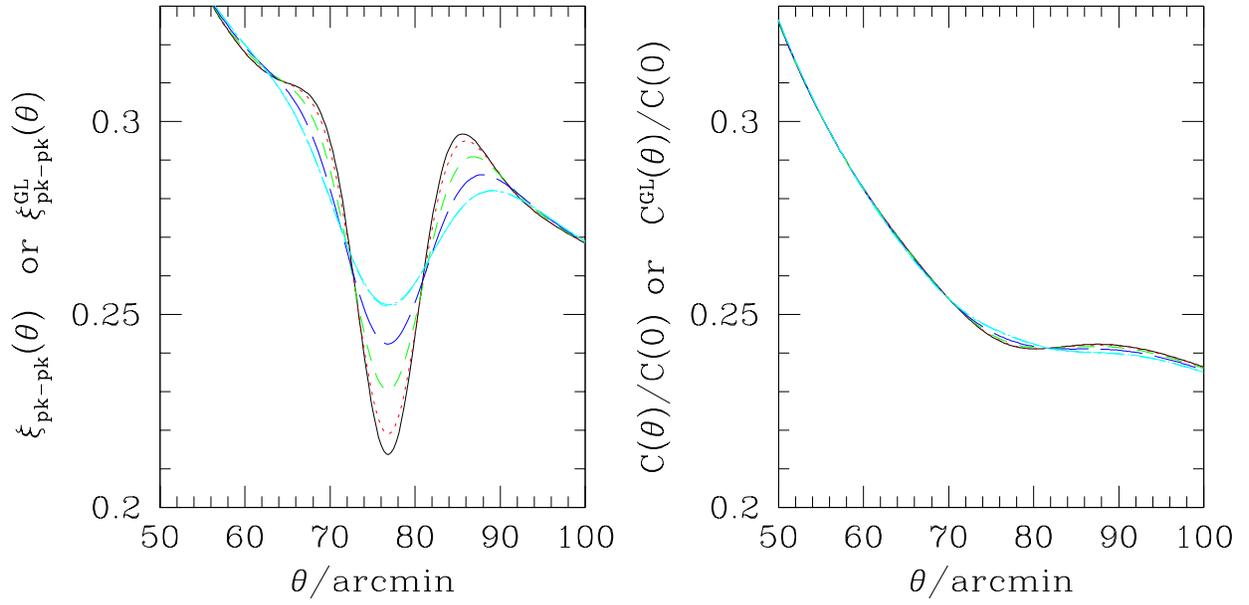}
 \end{center}
\figcaption{The two-point correlation functions of hotspots
  and the temperature anisotropies field around the 1st Doppler peak
  with and without the gravitational lensing effect.
\label{fig:compxi}}
\end{figure}

In Fig. \ref{fig:scdmxi}, we show the two-pint 
correlation function of hotspots of height 
above the threshold $\nu=1$ with and without gravitational lensing effect, 
$\xi^{\rm GL}_{\rm pk-pk}(\theta)$ and $\xi_{\rm pk-pk}(\theta)$,
respectively. 
Note that the intrinsic $\xi_{\rm pk-pk}(\theta)$ can be calculated by
using only $C_l$. The mean number density of hotspots is then
$\bar{n}_{\rm pk}(>\nu)=1.16\times 10^4 {\rm rad}^{-2}$. 
The dependence
of the lensing effect on the normalization of mass fluctuations
is demonstrated by the four choices of $\sigma_8=0.5,1,1.5,2$, where
$\sigma_8=0.5$ and $1$ roughly correspond to normalizations of
the cluster abundance (\cite{Eke}; \cite{Kitayama}) and
the COBE measurements (\cite{Bunn}), respectively. 
The figure clearly shows that $\xi_{\rm pk-pk}(\theta)$ has more 
oscillatory features than $C(\theta)$. This is because the gradient and second
derivative fields of $\Delta(\bm{\theta})$ used in the derivation of
$\xi_{\rm pk-pk}(\theta)$ have power spectra of $C_l l^4/(2\pi)$ and
$C_l l^6/(2\pi)$ per logarithmic interval in $l$, respectively, and thus
they strongly enhance the oscillations of $l^2C_l/(2\pi)$, while
$C(\theta)$ is fully determined by $l^2C_l/(2\pi)$. If
using the relation $l\approx\pi/\theta$, there are indeed 
correspondences between both oscillations of $\xi_{\rm pk-pk}(\theta)$ 
and $l^2C_l$ such as the depression at $\theta\approx 75'$
corresponding to the scales around the 1st Doppler peak of $C_l$,
a prominent peak at $\theta\approx 13'$, and 
a damping tail at $\theta<10'$ associated with the Silk
damping. Moreover, Fig. \ref{fig:scdmxi} shows
that the gravitational lensing
appears as a larger smoothing at scales where $\xi_{\rm 
pk-pk}(\theta)$ has more oscillatory features, since pairs of hotspots 
separated with the characteristic angular scales corresponding to the
oscillations of $\xi_{\rm pk-pk}$ are redistributed by the weak lensing 
and thus power of $\xi_{\rm pk-pk}(\theta)$ is transfered
from (or to) the scale to (or from) nearby scales.
Especially, it should be stressed that
even the lensing effect on the depression at $\theta\approx 75'$ associated with 
the 1st Doppler peak of $C_l$ is distinguishable where
the deviation $\delta \xi\ (\equiv (\xi^{\rm GL}_{\rm pk-pk}-\xi_{\rm 
pk-pk})/\xi_{\rm pk-pk})$ reaches up to $\approx 20\%$ for $\sigma_8=2.0$
while $\delta C<1.0\%$ in all cases (see Fig. \ref{fig:compxi}).
Since both $\xi^{\rm GL}_{\rm pk-pk}$ and $\xi_{\rm
pk-pk}$ give the same results as the shape becomes flat on larger
scales ($\theta> 80'$), the large scale amplitude of $\xi_{\rm pk-pk}$
gives the normalization independent of the lensing contributions. 
We also demonstrate the nonlinear corrections to the matter power spectrum
to clarify how the nonlinear evolution of matter fluctuations affects
our analysis. Using the semi-analytic formulae developed by Peacock \&
Dodds (1996) for the extreme case of $\sigma_8=2$, where the 
universe has too large power of the present-day small-scale matter
fluctuations, we found that the nonlinear
correction is small at $\theta>5'$.  
Most importantly, Fig. \ref{fig:scdmxi} indicates that
the magnitude of the lensing contributions to $\xi_{\rm pk-pk}$ is 
fairly sensitive to the amplitude of $\sigma_8$.
Furthermore, measuring scale-dependence of the lensing contributions
such as $\delta \xi(\theta)$ at $\theta=2', 16', 36',  44', 75'$ will allow
us to reconstruct the shape of projected power spectrum of matter fluctuations
in the range of $3h^{-1}\mbox{Mpc}\simlt \lambda\simlt 50h^{-1}\mbox{Mpc}$.

\section{Discussions and Conclusions}
In this Letter, we have investigated the weak lensing effect on the
two-point correlation of hotspots in the CMB.
One of the great advantages in studying the lensing effect on the CMB
anisotropies is 
that statistical properties of unlensed CMB field are fully predictable
by the Gaussian random theory. For example,
Bernardeau (1998) have investigated how the lensing
changes probability distributions of the ellipticity defined from the
local temperature curvature matrix, and shown that
the method can be a statistical indicator sensitive to the lensing
signatures.

It is now expected that the weak lensing can be a powerful probe
for measuring the power spectrum of matter fluctuations.
In case of the normalization based on measurements of the large angular
scale CMB anisotropies such as COBE normalization (e.g., \cite{Bunn}), 
we need to employ the very accurate matter transfer function which highly depends
on a lot of cosmological parameters in order to determine the amplitude of
small scale matter fluctuations. 
In fact, even most sensitive future experiments such as the
{\em Planck Surveyor} (\cite{Planck}) will constrain $\sigma_8$ up to
only $20\%$ accuracy (see, e.g., \cite{Bond}).
We thus wish to propose that $\xi_{\rm pk-pk}(\theta)$ can be
a new effective statistical tool for measuring 
amplitude and shape of the matter power spectrum.

For this purpose, we have to perform quantitative predictions whether or 
not the future satellite missions, {\em MAP} (\cite{MAP}) and {\em
Planck}, can measure
the lensing contributions to $\xi_{\rm pk-pk}(\theta)$ and constrain
$\sigma_8$ with high accuracy. In particular, we have to 
estimate not only uncertainties by the cosmic variance but also 
observational errors in identifying angular positions of hotspots in
a realistic 2D sky map of the CMB anisotropies.
It is then important to take into
account the following experimental effects. (1) The finite
beam size of detectors may incorporate some hotspots within
the beam size resulting from smoothing of the lensed CMB anisotropies
filed. The effect hence decreases the mean number density of hotspots 
and suppresses  two-point correlations of hotspots on scales below the beam size.
(2) The detector noise may make spurious hotspots in the observed
CMB map. HS estimated the errors
for measurements of the unlensed hotspots correlation functions expected 
from {\em MAP} and {\em Planck} by using numerical sky
realizations, and indeed showed that the theoretical predictions
including the above experimental effects
are in good agreement with the numerical results. 
Following their method, we are now investigating the detectability of
the lensing signatures to $\xi_{\rm pk-pk}(\theta)$,
and have obtained the preliminary result as follows. 
Supposing the data produced by
the {\em Planck} specification with $65\%$ sky coverage, 
the expected signal to noise ratios for the lensing signatures
are $S/N\approx 1.3, 1.8, 1.5$ and $ 3.1$
at $\theta=16', 36', 44'$ and $ 75'$,
respectively, for $\sigma_8=1.5$ model.
Furthermore, we expect that the statistical significance of $S/N$
could be increased by combining measurements of
the two-point correlations of {\em coldspots}.
The observed correlations of coldspots should be identical to that of
hotspots since they are subject to the Gaussian random statistics. 
These works are now in progress 
and will be presented in detail elsewhere.

We finally comment on 
possibilities that the lensing effect on $\xi_{\rm pk-pk}$ breaks
the so-called {\em geometrical degeneracy} (\cite{Bond};
\cite{Metcalf98}; \cite{Stompor}) inherent
in cosmological parameter determinations from
measurements of the CMB anisotropies. We have confirmed that the global
shape of $\xi_{\rm pk-pk}(\theta)$ is sensitive to fundamental
cosmological parameters such as $h,\Omega_{\rm b},\Omega_{\rm m0}$ and so on.
We thus expect that detecting $\theta$-dependence of the lensing
contributions to $\xi^{\rm GL}_{\rm pk-pk}(\theta)$ can yield an
additional constraint on the $\sigma_8-\Omega_{\rm m0}$ plane if we assume
the shape of matter power spectrum. Hence,
the degeneracy could be broken by observations of the CMB anisotropies
alone, without help of other astronomical measurements.

We would like to thank Takashi Murayama, Makoto Hattori, and David
N. Spergel for valuable comments and useful discussions.
We also would like to acknowledge Matias Zaldarriaga and Uros
Seljak for making available their CMBFAST code.
E.K. acknowledges support from a Japan Society for the Promotion of
Science fellowship.




\end{document}